\documentclass[preprint,superscriptaddress]{revtex4-1}
\usepackage{bibentry,natbib}
\usepackage{latexsym,bm}
\usepackage{graphicx}
\usepackage{rotating}
\usepackage{amsmath}
\usepackage{rotating}
\usepackage{color}
\usepackage[colorlinks,citecolor=red]{hyperref}

\DeclareGraphicsExtensions{.jpg}

\begin{document}
\title{Universal Three-Body Parameter of Heavy-Heavy-Light Systems with a Negative Intraspecies Scattering Length}
\author{Cai-Yun Zhao}
\affiliation{State Key Laboratory of Magnetic Resonance and Atomic and Molecular Physics, Wuhan Institute of Physics and Mathematics, Chinese Academy of Sciences, Wuhan 430071, P. R. China}
\affiliation{ University of Chinese Academy of Sciences,100088, Beijing, P. R. China}
\author{Hui-Li Han}
\email{huilihan@wipm.ac.cn}
\affiliation{State Key Laboratory of Magnetic Resonance and Atomic and Molecular Physics, Wuhan Institute of Physics and Mathematics, Chinese Academy of Sciences, Wuhan 430071, P. R. China}
\author{Meng-Shan Wu}
\affiliation{State Key Laboratory of Magnetic Resonance and Atomic and Molecular Physics, Wuhan Institute of Physics and Mathematics, Chinese Academy of Sciences, Wuhan 430071, P. R. China}
\author{Ting-Yun Shi}
\affiliation{State Key Laboratory of Magnetic Resonance and Atomic and Molecular Physics, Wuhan Institute of Physics and Mathematics, Chinese Academy of Sciences, Wuhan 430071, P. R. China}
\date{\today}

\begin{abstract}
The three-body parameter (3BP) $a^{\scriptscriptstyle(1)}_{\scriptscriptstyle-}$ is crucial for understanding Efimov physics, and a universal 3BP has been shown in experiments and theory in ultracold homonuclear gases. The 3BP of heteronuclear systems has been predicted to possess much richer properties than its homonuclear counterparts due to the larger parameter space. In this work, we investigate the universal properties of $a^{\scriptscriptstyle(1)}_{\scriptscriptstyle-}$ for heavy-heavy-light (HHL) systems with a negative intraspecies scattering length $a_{\scriptscriptstyle\textsl{HH}}$. We find that $a^{\scriptscriptstyle(1)}_{\scriptscriptstyle-}$ follows a universal behavior determined by the van der Waals (vdW) interaction and the mass ratio. A linear fit of $a^{\scriptscriptstyle(1)}_{\scriptscriptstyle-}$ for large $a_{\scriptscriptstyle\textsl{HH}}$ leads to a simple formula, which gives the general dependence of $a^{\scriptscriptstyle(1)}_{\scriptscriptstyle-}$ on the mass ratio and the vdW length $r_{\scriptscriptstyle\textsl{vdW,HL}}$. In a special case, when the two heavy atoms are in resonance, $a^{\scriptscriptstyle(1)}_{\scriptscriptstyle-}$ is approximately constant: $a^{\scriptscriptstyle(1)}_{\scriptscriptstyle-} = -(6.3\pm0.6)\, r_{\scriptscriptstyle\textsl{vdW,HL}}$.
\end{abstract}
\maketitle
\section{Introduction}
The Efimov effect, which was predicted
by Efimov in 1970\;\cite{Efimov1971,Efimov1973}, has attracted both experimental\;\cite{Kraemer2006Cs,Zaccanti2009Cs,Lompe2010Li6,Berninger2011Cs,Huang2014Cs,
Williams2009Li6,Huckans2009Li6,Ottenstein2008Li6,Pollack2009Li7,Gross2009Li7,Gross2010Li7,
Kunitski2015He,
Barontini2009rb87k41,Bloom2013rb87k40,HuMingGuang2014rb87k40,Wacker2016rb87k3941,Kato2017rb87k4041,Maier2015rb87Li7,Pires2014cslia-,
Tung2014cslia-,Ulmanis2016cslia+,Ulmanis_Kuhnle2016csli,Johansen2017csli}
and theoretical\;\cite{jiawangPRL2012,yujunwangPRl2012,Sorensen2012,Wumengshan2014,Huang2014Li6,Wumengshan2016,Hafner2017,Hanhuili2018,Wenz2009Li6}
attention due to its novel phenomena, such as universality and discrete scale invariance. The  most dramatic manifestation of the Efimov effect is the possibility of infinite three-body bound states when the two-body scattering length $a$ is large compared to the characteristic range $r_0$ of the two-body interaction potential\;\cite{Efimov1971}. This infinity of trimer states follows a discrete symmetry scaling, i.e.,\;$E_n=\lambda^2E_{n+1}$, where $\lambda=e^{\pi/ s_0}$ is the scaling constant and $s_0$ is a universal parameter depending on the number of resonant interactions, the quantum statistics and the mass ratio of the trimer's constituents\;\cite{hwhammer2006,efimov_phisics,D'Incao2006mass}.

In addition to their discrete scaling property, Efimov states are characterized by the three-body parameter (3BP), which fully determines the Efimov spectrum in the Efimov window of universality. The 3BP $a_{\scriptscriptstyle-} = a_{\scriptscriptstyle-}^{\scriptscriptstyle(n)}/\lambda^{\scriptscriptstyle n-1}$ is defined, for sufficiently large $n$, by the scattering length $a_{\scriptscriptstyle-}^{\scriptscriptstyle(n)}$ at which the $n$th excited Efimov bound state emerges from the three-body continum. In a realistic system, it is often difficult to access the universal window corresponding to highly excited trimer state, and thus the scattering length $a_{\scriptscriptstyle-}^{\scriptscriptstyle(1)}$ at which the ground trimer state appears constitues an approximation of the true 3BP\;\cite{hwhammer2006,efimov_phisics}. For homonuclear systems, experiments with
alkali-metal atoms have observed a universal value for $a_{\scriptscriptstyle-}^{\scriptscriptstyle(1)}$ when expressed in terms of the van der Waals (vdW) length $r_{\scriptscriptstyle\textsl{vdW}}$: $a_{\scriptscriptstyle-}^{\scriptscriptstyle (1)} = -(8.9 \pm1.8) r_{\scriptscriptstyle\textsl{vdW}}$\;$[r_{\scriptscriptstyle\textsl{vdW}} = (2 \mu_{\scriptscriptstyle 2b} C_{6})^{1/4}/2]$\;\cite{Ottenstein2008Li6,Huckans2009Li6,Williams2009Li6,Pollack2009Li7,Gross2009Li7,Zaccanti2009Cs,
Lompe2010Li6,Gross2010Li7,Berninger2011Cs,Huang2014Cs,Wild2012Rb85,Roy2013K39,chinchen}.
 Calculations from theories based on a single-channel van der Waals model confirmed this vdW universality for broad Feshbach resonances\;\cite{yujunwangPRl2012,jiawangPRL2012,Huang2014Li6,Naidon2014PRL,Naidon2014PRLA}. In recent studies, the observed nonuniversal Efimov ground state locations further confirm that the vdW universality of 3BP is valid only for broad Feshbach resonances\;\cite{Chapurin2019,PhysRevA.97.033623,Johansen2017csli}.
 For heteronuclear systems, i.e. heavy-heavy-light (HHL) systems, the system possesses the mass ratio $M/m$, two scattering lengths ( intra- and interspecies scattering lengths $a_{\scriptscriptstyle\textsl{HH}}$, $a_{\scriptscriptstyle\textsl{HL}}$ ), and two different vdW lengths (the H-H and H-L vdW lengths:  $r_{\scriptscriptstyle\textsl{vdW,HH/HL}} = ( 2\mu_{\scriptscriptstyle 2b} C_{\scriptscriptstyle 6, HH/HL})^{1/4}/2 $ ). Thus, the universal property of 3BP is significantly more complicated due to the larger parameter space. In the work of Ref.\;\cite{yujunwangPRl2012}, the authors predicted that the 3BP of the heteronuclear system depends on not only the pairwise vdW tails but also the mass ratio and the intraspecies scattering length $a_{\scriptscriptstyle\textsl{HH}}$ between two heavy atoms\;\cite{yujunwangPRl2012}. The dependence of the Efimov energy spectrum $E_{\scriptscriptstyle\textsl{n}}(a_{\scriptscriptstyle\textsl{HH}})$ at the heteronuclear unitary ($a_{\scriptscriptstyle\textsl{HL}}\rightarrow\infty$) was also reported in Ref\;\cite{yujunwangPRl2012}. Recent experimental evidence in the $^{133}$Cs\,-$^{133}$Cs\,-$^{6}$Li system suggests that the first Efimov resonance position depends on the sign of the Cs\,-Cs scattering length ($a_{\scriptscriptstyle\textsl{CsCs}}>0$ or $a_{\scriptscriptstyle \textsl{CsCs}}<0$, $a_{\scriptscriptstyle\textsl{AB}}$ represents the A-B scattering length throughout this paper)\;\cite{Pires2014cslia-,Tung2014cslia-,Ulmanis2016cslia+,Hafner2017,Ulmanis2016NJPcsli}. This nonuniversal phenomenon has led to further investigations regarding the effects of the intraspecies scattering length on the heteronuclear Efimov scenario\;\cite{Ulmanis2016cslia+,PhysRevLett.120.023401}. The study of Ref\;\cite{Ulmanis2016cslia+} indicated that there exists two Efimov branches for repulsive intraspecies interactions, and the Efimov states belonging to the lower branch will merge with the CsCs + Li threshold and thus can lead to an absence of Efimov resonances in three-body recombination spectra.

Compared to homonuclear systems, the available experimental data for Efimov features in heteronuclear systems are relatively sparse and consequently do not clearly display the universality of 3BP. To date, convincing Efimov resonances were observed only in $^{7}$Li-$^{87}$Rb\,\cite{Maier2015rb87Li7} and $^{6}$Li-$^{133}$Cs\;\cite{Pires2014cslia-,Tung2014cslia-,Ulmanis2016cslia+,Hafner2017,Ulmanis2016NJPcsli} mixtures, even though the existing theoretical works \;\cite{yujunwangPRl2012,Ulmanis2016cslia+,PhysRevLett.120.023401} have predicted some universal properties of 3BP in heteronuclear systems. There are still questions regarding how exactly the vdW interaction and intraspecies scattering length affect the universal behavior of $a_{\scriptscriptstyle-}^{\scriptscriptstyle(1)}$ and how the $a_{\scriptscriptstyle-}^{\scriptscriptstyle(1)}$ behaves across different HHL systems, which need to be understood deeply.

Here, we investigate the $a_{\scriptscriptstyle-}^{\scriptscriptstyle(1)}$ of ten experimentally interesting systems for the negative intraspecies scattering length case. Compared to the case of positive intraspecies scattering length, the $a_{\scriptscriptstyle-}^{\scriptscriptstyle(1)}$ of HHL systems with $a_{\scriptscriptstyle\textsl{HH}} < 0$ has been the subject of limited investigations both theoretically and experimentally. To date, the $a_{\scriptscriptstyle-}^{\scriptscriptstyle(1)}$ with negative intraspecies scattering length has been investigated only in $^{133}$Cs-$^{6}$Li mixtures \;\cite{Pires2014cslia-,Tung2014cslia-,Ulmanis2016cslia+,Hafner2017,Ulmanis2016NJPcsli}. Information about $a_{\scriptscriptstyle-}^{\scriptscriptstyle(1)}$ for other experimentally interesting systems with negative intraspecies scattering length is absent. Our intent is to give the value of $a_{\scriptscriptstyle-}^{\scriptscriptstyle(1)}$ for some experimentally interesting systems for a possible fixed value of the intraspecies scattering length and further extract its universal behavior with respect to the intraspecies scattering length and vdW interaction across many different HHL systems. Our study represents a further step toward understanding the universal behavior of $a_{\scriptscriptstyle-}^{\scriptscriptstyle(1)}$ in heteronuclear systems.

\section{theoretical formalism}
The three-body Schr$\mathrm{\ddot{o}}$dinger equation describing the relative motion of a system can be written as
\begin{equation}
\label{schrodinger}
[-\frac{1}{M}\nabla^2_\textbf{r}-\frac{2M+m}{4Mm}\nabla^2_{\boldsymbol{\rho}}
+V_{\scriptscriptstyle\textsl{HH}}(r)+V_{\scriptscriptstyle\textsl{ HL}}(|\boldsymbol{\rho}+\frac{\textbf{r}}{2}|)
+V_{\scriptscriptstyle\textsl{HL}}(|\boldsymbol{\rho}-\frac{\textbf{r}}{2}|)]\Psi =E\Psi\,,
\end{equation}
where $ \textbf{r}$ is the displacement vector between two heavy Boson atoms with mass $M$, and $\boldsymbol{\rho}$ is the relative position between their center of mass and the light atom $m$\;\cite{hwhammer2006,efimov_phisics}.
We adopt the Lennard-Jones potential with a vdW tail to describe the interaction between atoms with a distance $r_{\scriptscriptstyle\textsl{{HH/HL}}}$:
\begin{align}
\label{eq:ev2}
V_{\scriptscriptstyle\textsl{HH/HL}}(r_{\scriptscriptstyle\textsl{HH/HL}})
=-\frac{C_{\scriptscriptstyle\textsl{6,HH/HL}}}
{r^{\scriptscriptstyle6}_{\scriptscriptstyle\textsl{HH/HL}}}
[1-\frac{1}{2}(\frac{\lambda_{\scriptscriptstyle\textsl{HH/HL}}}
{r_{\scriptscriptstyle\textsl{HH/HL}}})^6]\,,
\end{align}
where $\lambda_{\scriptscriptstyle\textsl{HH/HL}}$ is adjusted to give the desired scattering length $a_{\scriptscriptstyle\textsl{HH/HL}}$ and $C_{6,\scriptscriptstyle\textsl{HH/HL}}$ is the usual dispersion coefficient. The Lennard-Jones potential has been demonstrated to be an excellent model potential for studying the universality of 3BP\;\cite{jiawangPRL2012,yujunwangPRl2012}.

In the frame of hyperspherical coordinates, the hyperradius $R$ of the following form can describe the global size:
\begin{align}
\label{eq:R}
\mu R^2=\frac{M}{2}r^2+\frac{2Mm}{2M+m}\rho^2\,,
\end{align}
where $\mu$ is the three-body reduced mass\;\cite{CDLIN1995}.

The Schr\"{o}dinger equation\.(\ref{schrodinger})can be written in the hyperspherical coordinates with the rescaled
wave functions $\psi=\Psi R^{5/2}$ as follows:

 \begin{equation}
\bigg[-\frac{1}{2\mu}\frac{d^{2}}{dR^{2}}+\bigg(\frac{\Lambda^{2}-\frac{1}{4}}{2\mu R^{2}}+
V_{\scriptscriptstyle\textsl{HH}}(r_{\scriptscriptstyle\textsl{HH}}) + V_{\scriptscriptstyle\textsl{HL}}(r_{\scriptscriptstyle\textsl{HL}}) + V_{\scriptscriptstyle\textsl{HL}}(r_{\scriptscriptstyle\textsl{HL}})\bigg)\bigg]\psi =E\psi\,,
\label{eq:s4}
 \end{equation}

where $\Lambda^{2}$ is the squared ``grand angular momentum operator", whose expression is given in Ref.\;\cite{CDLIN1995}.

We first determine the adiabatic potentials $U_{\nu}(R)$ and the corresponding channel
functions $\Phi_{\nu}(R,\Omega)$ [ $ \Omega\equiv(\theta,\phi,\alpha,\beta,\gamma)$ is the hyperangle ] at the fixed $R$ by solving the following adiabatic eigenvalue equation:
\begin{equation}
\bigg(\frac{\Lambda^{2}-\frac{1}{4}}{2\mu R^{2}}+
V_{\scriptscriptstyle\textsl{HH}}(r_{\scriptscriptstyle\textsl{HH}}) + V_{\scriptscriptstyle\textsl{HL}}(r_{\scriptscriptstyle\textsl{HL}}) + V_{\scriptscriptstyle\textsl{HL}}(r_{\scriptscriptstyle\textsl{HL}})\bigg)\Phi_{\nu}(R,\Omega)=U_{\nu}(R)\Phi_{\nu}(R,\Omega)\,.
\label{eq:s7}
\end{equation}

The wave function $\psi$ can then be expanded with the complete, orthonormal adiabatic channel functions $ \Phi_\upsilon$ by
\begin{align}
\label{eq:wf}
 \psi(R,\Omega)=\sum^\infty_{\nu=0}F_\nu(R)\Phi_\nu(R,\Omega)\,,
\end{align}
and substitute $\psi(R,\Omega)$ into Eq.~(\ref{eq:s4}) giving a set of coupled ordinary differential equations:
\begin{align}
[-\frac{1}{2\mu}\frac{d^2}{dR^2}+U_\nu(R)- E]F_{\nu,\scriptscriptstyle E}(R)
-\frac{1}{2\mu}\sum_{\nu'}[2P_{\nu\nu'}(R)\frac{d}{dR}+Q_{\nu\nu'}(R)]F_{\nu',\scriptscriptstyle E}(R)=0\,,
\end{align}

where
\begin{align}
P_{\nu\nu'}=\int d\Omega \Phi_{\nu}(R;\Omega)^{*}\frac{\partial}{\partial R}\Phi_{\nu'}(R;\Omega),
\end{align}
and
\begin{align}
Q_{\nu\nu'} = \int d\Omega \phi_{\nu}(R;\Omega)^{*}\frac{\partial^{2}}{\partial R^{2}}\Phi_{\nu'}(R;\Omega)
\end{align}
are the nonadiabatic couplings.
\section{ results and discussion}
\subsection{The first Efimov resonance position $a_{\scriptscriptstyle-}^{\scriptscriptstyle(1)}$ for some HHL systems}
Table\;\ref{t1} summarizes our results obtained for the $a_{\scriptscriptstyle-}^{\scriptscriptstyle(1)}$ relevant to the three-body recombination process, $H + H + L \rightarrow HL + H$, for some HHL mixtures. The $a_{\scriptscriptstyle-}^{\scriptscriptstyle(1)}$ corresponds to the interspecies scattering length $a_{\scriptscriptstyle\textsl{HL}}$ at which an Efimov state reaches the three-body threshold. Experimentally, the inter- and intraspecies interactions are generally not controlled independently and therefore will generate background intraspecies scattering length $a_{\scriptscriptstyle\textsl{HH}}$ when tuning the $H-L$ interaction. We give the value of $a_{\scriptscriptstyle-}^{\scriptscriptstyle(1)}$ near a $ H + L $ Feshbach resonance and also the range of it at the given value of the intraspecies scattering length $a_{\scriptscriptstyle\textsl{HH}}$. For comparison, we also show the results for isotopic systems that have the value of $a_{\scriptscriptstyle-}^{\scriptscriptstyle(1)}$ for positive intraspecies scattering length $a_{\scriptscriptstyle\textsl{HH}} > 0$.

We calculate $a_{\scriptscriptstyle-}^{\scriptscriptstyle(1)}$ for the $^{133}$Cs\,-$^{133}$Cs\,-$^{6}$Li system and compare it with the experimental results to ensure the reliability of our method. With $a_{\scriptscriptstyle\textsl{CsCs}}=-1\,200\,a_0$, the calculated $a_{\scriptscriptstyle-}^{\scriptscriptstyle(1)}$ is $-313\, a_0$, which agrees with the experimental result $-311(3)\, a_0$\cite{Pires2014cslia-,Ulmanis2016NJPcsli,Hafner2017} and also the measurement of Ref \cite{Tung2014cslia-}, as indicated in Table\;\ref{t1}.

In addition to the $^{133}$Cs\,-$^{133}$Cs\,-$^{6}$Li system, the Rb\,-Li heteronuclear mixture is another strong mass-imbalanced system that is suitable for studying the universality of Efimov physics. The first Efimov resonance was observed at $-1\,870(121)\,a_0$ for the $^{87}$Rb\,-$^{87}$Rb\,-$^{7}$Li system\;\cite{Maier2015rb87Li7}. Based on the known scaling factor, a second Efimov resonance is expected at about $-15\,000\,a_{0} $ and a third resonance at $-115\,000\,a_{0}$ \;\cite{Maier2015rb87Li7}. The analyses of Ref.\;\cite{Maier2015rb87Li7} show that it may be difficult to observe three Efimov features in $^{87}$Rb\,-$^{7}$Li mixture with current experiment conditions. The $^{85}$Rb\,-$^{6}$Li system has a broad Feshbach resonance at the magnetic field $B=40\,G$\;\cite{feshbach_rb85li6}, which leads to a $a_{\scriptscriptstyle\textsl{$^{85}$Rb$^{85}$Rb  }}=-450\,a_{0} $ background scattering length in the $^{85}$Rb\,-$^{6}$Li mixture. The calculated first Efimov resonance is found at $a_{-}^{\scriptscriptstyle(1)} = -337\,a_{0}$ for the $^{85}$Rb\,-$^{85}$Rb\,-$^{6}$Li system. In this case, the second
Efimov resonance is expected to be at about $-2\,346\, a_{0}$ and the third resonance at $-16\,334\,a_{0}$. Therefore, observing three Efimov features in the $^{85}$Rb\,-$^{85}$Rb\,-$^{6}$Li system seems to be within experimental reach.

  Heteronuclear Efimov states have been searched for in $^{87}$Rb\,-$^{87}$Rb\,-$^{41,40,39}$K since 2009\cite{Barontini2009rb87k41,Bloom2013rb87k40,HuMingGuang2014rb87k40,Wacker2016rb87k3941,Kato2017rb87k4041}
. The first Efimov resonance is expected to occur around the Rb\,-K scattering length $a_{-}^{\scriptscriptstyle(1)}\leq -30\,000\,a_{0}$ due to the positive background scattering length $a_{\scriptscriptstyle\textsl{$^{87}$Rb$^{87}$Rb}}\approx100\,a_0$\;\cite{feshbach_rb87rb87}. The very large interspecies scattering length is outside the current experiment reach. Thus, no convincing Efimov resonance has been observed in these systems to date. Fortunately, $^{85}$Rb has negative scattering lengths in the field range between $0 $ and $1\,000\, G$~\cite{feshbach_rb85rb85}, and both $^{85}$Rb\,-$^{41}$K\;\cite{feshbach_rb85k41} and $^{85}$Rb\,-$^{40}$K\;\cite{feshbach_alkali-metal} have magnetic Feshbach resonances in this range. Near the $B = 339\, G$ Feshbach resonance of $^{85}$Rb\,-$^{40}$K\;\cite{feshbach_alkali-metal}, the $^{85}$Rb\,-$^{85}$Rb\,-$^{40}$K system has a background scattering length $a_{\scriptscriptstyle\textsl{$^{85}$Rb$^{85}$Rb}}=-416\,a_{0} $. The calculated first Efimov resonance is found at $a_{\scriptscriptstyle-}^{\scriptscriptstyle(1)} = -1\,036\,a_{0}$ in this case, which is far smaller than that in the positive background scattering case and may be observed experimentally.

 The $^{85}$Rb\;-$^{85}$Rb\,-$^{23}$Na is also a less mass-imbalanced system whose mass ratio is $3.7$, as shown in Table\;\ref{t2}. Its isotopic system $^{87}$Rb\,-$^{23}$Na mixtures have been investigated by combined experimental-theoretical study by exploring its three- and four-body processes\,\cite{dajunWang2016Rb87Na23}. The present calculated value of $a_{\scriptscriptstyle-}^{\scriptscriptstyle(1)}$ is $-685\,a_{0}$ for $^{85}$Rb\;-$^{85}$Rb\,-$^{23}$Na near the $B = 314\, G$ Feshbach resonance of $^{85}$Rb\,-$^{23}$Na. Reference\,\cite{PhysRevA.100.042706} has predicted the
 $a_{\scriptscriptstyle-}^{\scriptscriptstyle(1)} = - 11\,850\, a_{0}$ for its isotopic system $^{87}$Rb\;-$^{87}$Rb\,-$^{23}$Na using the finite-range model.

 The above systems are examples whose $a_{\scriptscriptstyle-}^{\scriptscriptstyle(1)}$ for negative intraspecies scattering length is much larger than the value of positive intraspecies scattering length. Our calculations, together with the existing data for positive intraspecies scattering length, demonstrated that systems with negative intraspecies scattering lengths are more suitable for observing the Efimov ground state than those with positive scattering lengths. This result is consistent with the implication of the experimental results of the $^{133}$Cs\,-$^{133}$Cs\,-$^{6}$Li system with different signs of the intraspecies scattering length\;\cite{Pires2014cslia-,Tung2014cslia-,Ulmanis2016cslia+,Hafner2017,Ulmanis2016NJPcsli}.

$^{85}$Rb\,-He$^{*}$(2 $^{3}$S) and $^{176}$Yb\,-Li mixtures both have large mass ratios and thus are the promising systems to observe multi-Efimov features. There are no Feshbach resonances of these systems reported in the literature. We provide the values of $a_{\scriptscriptstyle-}^{\scriptscriptstyle(1)}$ in $^{85}$Rb\,-$^{85}$Rb\,-$^{3}$He$^*(2^3S)$ and $^{176}$Yb\,-$^{176}$Yb\,-$^{6}$Li systems for the possible range of the background scattering length, as listed in Table\;\ref{t1}. It is important to note that, the information of $a_{\scriptscriptstyle-}^{\scriptscriptstyle(1)}$ permits more precise of Efimov resonance position in ultracold quantum gases, providing benefits to experiments for more accurate measurements.
\begin{table}
\caption{\label{t1} Calculated and observed Efimov resonance position $a_{\scriptscriptstyle-}^{\scriptscriptstyle(1)}$ with different background scattering lengths $a_{\scriptscriptstyle \textsl{HH}}$ and experimental magnetic fields $B_{\scriptscriptstyle \textsl{HL}}(G)$ of the H-L Feshbach resonance. The universal Efimov scaling constants $s_0$ and $s_0^*$ correspond to two resonant interactions and three resonant interactions, respectively. For comparison, we also list the results of the isotopic system with positive intraspecies scattering length.
The experimental results are shown in bold.   }
\renewcommand\arraystretch{0.17}
\begin{tabular*}{16.5cm}{@{\extracolsep{\fill}}lcccccc}
   \hline
   \hline

$system( {\scriptstyle \textsl{H-H-L}})$ &$B_{\scriptscriptstyle\textsl{HL}}(G)$ &$a_{\scriptscriptstyle \textsl{HH}}(a_0)$  &\multicolumn{2}{c}{$a^{\scriptscriptstyle(1)}_{\scriptscriptstyle-}(a_0)$}  &$s_0$        &$s_0^*$  \\
\cline{4-5}
&& &this\;work & Ref. &       & \\
\hline
$^{133}$Cs\,-$^{133}$Cs\,-$^{6}$Li  &$\textbf{849}\,^a$    &$\textbf{-1\,200}\,^a$  &$-313$    & $\textbf{311(3)}\,^a$       &$1.983$&$2.003$\\
                                    &$\textbf{848.55}\,^b$ &$\textbf{-1\,240}\,^b $ &          & $\textbf{-323(8)}\,^b$        &       &     \\
$               $                   &$ $                   &$-448\sim-10^6$         &$-351\sim-285$                &         &       &  \\
                                    &$\textbf{889}\,^c$    &$\textbf{190}\,^c$      &          &$\textbf{-2\,150(50)}\,^{c}$   &       &  \\
                                    &                      &                        &                              &  &       &\\
\hline
$^{85}$Rb\,-$^{85}$Rb\,-$^{7}$Li    & $143\,^d $&$-430\,^e$           &$-363 $                       &  &$1.509$&$1.575$\\
$               $                   & $     $          &$-400\sim-1\,990$           &$-367\sim-300$                &  &       &  \\
$               $                   & $     $          &$-1\,990\sim-10^6$          &$-300\sim-279$                &  &       &  \\
$^{85}$Rb\,-$^{85}$Rb\,-$^{6}$Li    & $40\,^f$  &$-450\,^e$           &$-337 $                       &  &$1.619$&$1.670$\\
$               $                   & $     $          &$-400\sim-1\,990$           &$-343\sim-284$                &  &       &  \\
$               $                   & $     $          &$-1\,990\sim-10^6$          &$-284\sim-265$                &  &       &  \\
$^{87}$Rb\,-$^{87}$Rb\,-$^{7}$Li    &$\textbf{661}\,^h$&$100\,^g\;\;\;$      &     &$\textbf{-1870(121)}\,^h$  &$1.521$&$1.585$\\
                                    & $ $              &$         $              &             & &       &\\
\hline
$^{85}$Rb\,-$^{85}$Rb\,-$^{41}$K    &$180 \sim 190\,^i$   &$-430\,^e$ &$-1\,041 $                  &  &$0.639$&$1.039$    \\
$               $                   &$656 \sim 681\,^i$  &$-400\,^e$ &$-1\,088 $   & &       &   \\
$               $                   & $     $          &$-400\sim-500$         &$-1\,088\sim-960$ &  &       &  \\
$               $                   & $     $          &$-500\sim-1\,990$      &$-960\sim-596 $ &  &       &  \\
$               $                   & $     $          &$-1\,990\sim-10^6$     &$-596\sim-478 $ &  &       &  \\
$^{85}$Rb\,-$^{85}$Rb\,-$^{40}$K    & $339\,^j$        &$-416\,^e$      &$-1036$                            &  &$0.647$&$1.041$   \\
$               $                   & $     $          &$-400\sim-500$         &$-1\,072\sim-946$&  &       &   \\
$               $                   & $     $          &$-500\sim-1\,990$      &$-946\sim-591 $ &  &       &  \\
$               $                   & $     $          &$-1\,990\sim-10^6$     &$-591\sim-477 $ &  &       &  \\
$^{87}$Rb\,-$^{87}$Rb\,-$^{40}$K    & $     $          &$100\,^g$      &    &$<-3\times10^4\,^k$  &$0.653$&$1.043$\\                                    &                                                      &                        &              &   &       &  \\
\hline
$^{85}$Rb\;-$^{85}$Rb\,-$^{23}$Na   & $314\,^l$        &$-410\,^e$   &$-685 $         &   &$0.863$&$1.127$\\
$               $                   & $     $          &$-400\sim-1\,990$       &$-691\sim-456 $ &  &       &  \\
$               $                   & $     $          &$-1\,990\sim-10^6$      &$-456\sim-390 $ &  &      & \\
$^{87}$Rb\;-$^{87}$Rb\,-$^{23}$Na   & $ $              &$100\,^g $      & &$-11\,850\,^m$  &$0.870$&$1.130$\\
                                    &                  &                        &              & &       &  \\
                                    \hline
$^{85}$Rb\,-$^{85}$Rb\,-$^{4}$He$^*(2^3S)$&            &$-400\sim-830$          &$-306\sim-280 $& &$1.952$&$1.974$  \\
$               $                   & $     $          &$-830\sim-10^6$         &$-280\sim-250 $& &       &  \\
                                    &                  &                        &               & &       &\\
$^{85}$Rb\,-$^{85}$Rb\,-$^{3}$He$^*(2^3S)$&            &$-400\sim-500$          &$-272\sim-265 $& &$2.225$&$2.236$  \\
$               $                   & $     $          &$-500\sim-10^6$         &$-265\sim-230 $& &       &  \\
                                    &                  &                        &               & &       &\\
$^{176}$Yb\,-$^{176}$Yb\,-$^{6}$Li  &                  &$-400\sim-10^6$\,$^n$     &$-257\sim-218 $& &$2.258$&$2.268$ \\
                                    &                  &                        &               & &       &  \\
\hline
\hline
\end{tabular*}
\begin{flushleft}
\vskip-1.5ex
$^{a}$From\cite{Pires2014cslia-,Ulmanis2016NJPcsli,Hafner2017} ,
$^{b}$From\cite{Tung2014cslia-},
$^{c}$From\cite{Ulmanis2016cslia+},
$^{d}$From\cite{feshbach_alkali-metal},
$^{e}$From\cite{feshbach_rb85rb85},
$^{f}$From\cite{feshbach_rb85li6},
$^{g}$From\cite{feshbach_rb87rb87},
$^{h}$From\cite{Maier2015rb87Li7},
$^{i}$From\cite{feshbach_rb85k41},
$^{j}$From\cite{feshbach_alkali-metal},
$^{k}$From\cite{yujunwangPRl2012},
$^{l}$From\cite{feshbach_Rb85Na23},
$^{m}$From\cite{dajunWang2016Rb87Na23},
$^{n}$From\cite{YbYbScatteringLength}
\end{flushleft}
\end{table}

\subsection{Dependence of $a_{\scriptscriptstyle-}^{\scriptscriptstyle(1)}$ on $ a_{\scriptscriptstyle\textsl{HH}}$ for different mass ratios }
\begin{figure}
 \centering
   \includegraphics[width=0.7\linewidth]{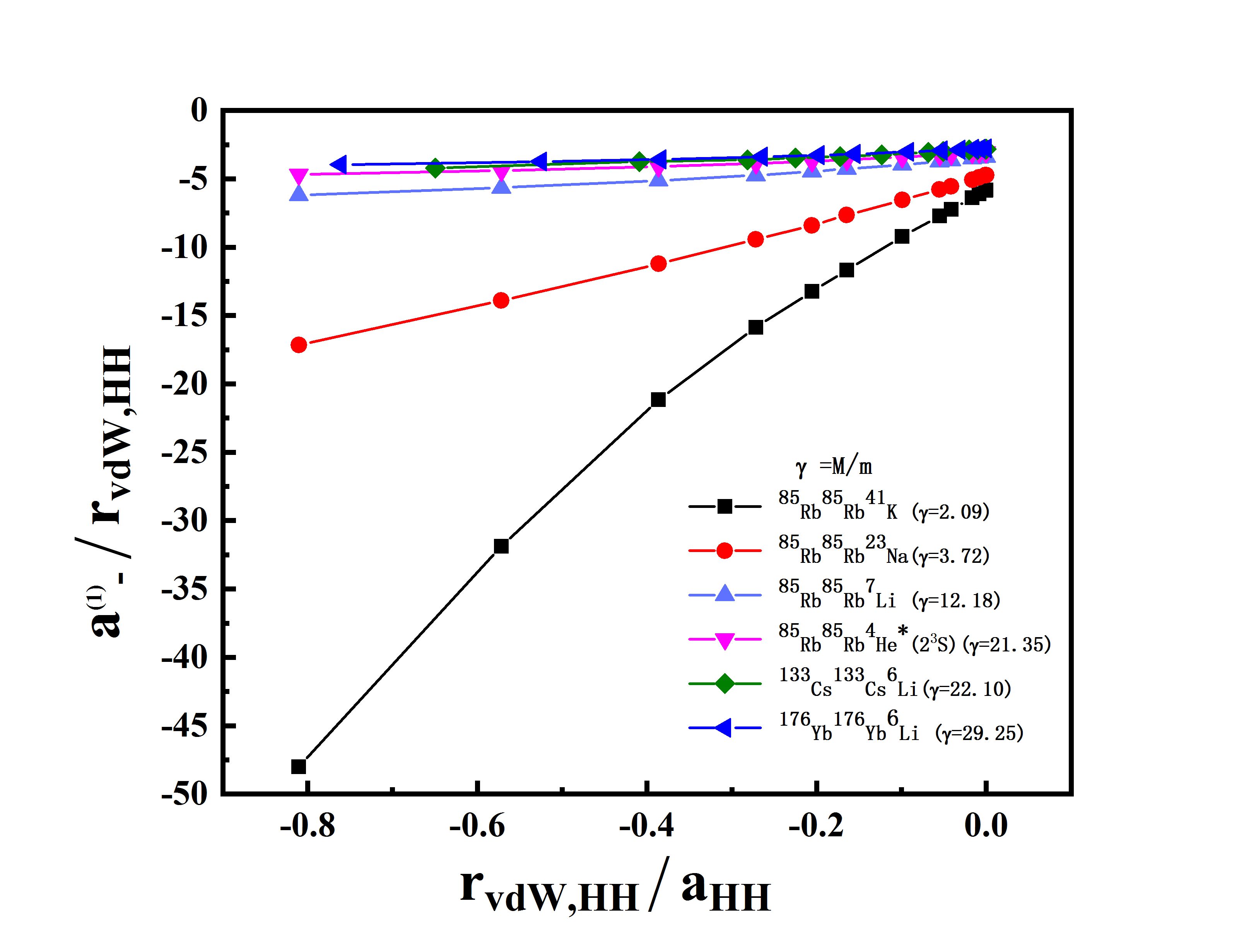}
   \caption{$a_{\scriptscriptstyle-}^{\scriptscriptstyle(1)}$ plotted vs the inverse intraspecies scattering length $1/a_{\scriptscriptstyle\textsl{HH}}$. All quantities are scaled by the intraspecies vdW length. }
   \label{HH1}
  \end{figure}

To investigate the dependence of $a_{\scriptscriptstyle-}^{\scriptscriptstyle(1)}$ on the mass ratio and the background scattering length, we plot $a_{\scriptscriptstyle-}^{\scriptscriptstyle(1)}$ as a function of $1/a_{\scriptscriptstyle\textsl{HH}}$ in Fig.\;\ref{HH1}. The intraspecies scattering length begins from about $- \, r_{\scriptscriptstyle\textsl{vdW,HH}}$ to infinite. It can be observed that $|a^{\scriptscriptstyle(1)}_{\scriptscriptstyle-}|$ decreases when the heavy-heavy interactions approach the unitary limit. For less mass-imbalanced systems such as $^{85}$\,Rb-$^{85}$Rb\,-$^{41,40}$K and $^{85}$Rb\,-$^{85}$Rb\,-$^{23}$Na, $a_{\scriptscriptstyle-}^{\scriptscriptstyle(1)}$ has a dramatically strong dependence on the background scattering length $a_{\scriptscriptstyle\textsl{HH}}$. However, for strong mass-imbalanced systems such as $^{133}$Cs\,-$^{133}$Cs\,-$^{6}$Li, $^{172}$Yb\,-$^{172}$Yb\,-$^{6}$Li and $^{85}$Rb\,-$^{85}$Rb\,-$^{4,3}$He$^*$(2$^3$S), the lines are flatter, which means the heavy-heavy interaction plays a less important role in determining the value of $a_{\scriptscriptstyle-}^{\scriptscriptstyle(1)}$. This is consistent with the fact that the Efimov scaling factor for a high mass ratio is insensitive to the heavy-heavy interaction\;\cite{PhysRevA}.
\subsection{The universal properties of $a_{\scriptscriptstyle-}^{\scriptscriptstyle(1)}$ for the HHL system}
\begin{figure}
 \centering
   \includegraphics[width=0.7\linewidth]{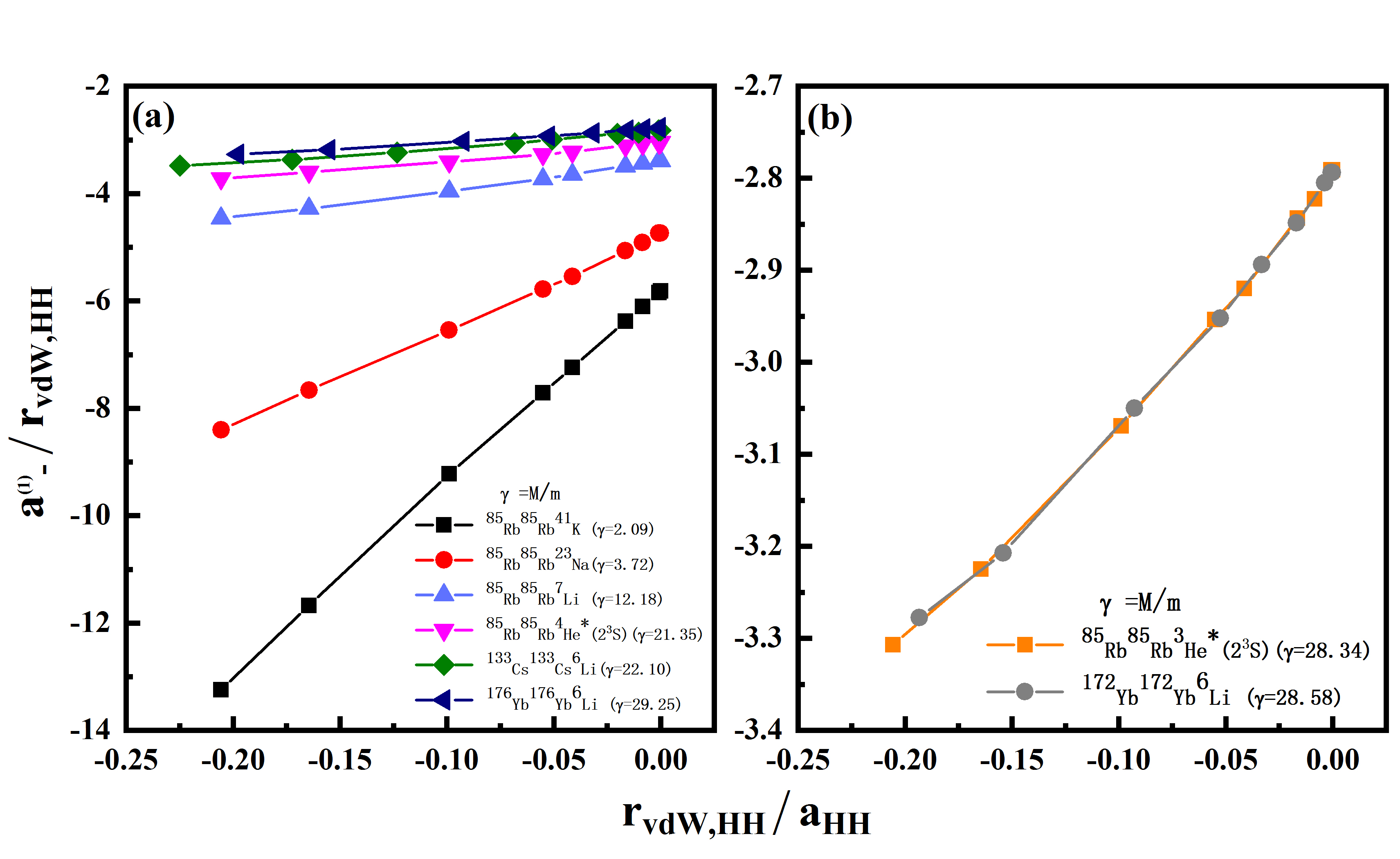}
   \caption{linear dependence of $a_{\scriptscriptstyle-}^{\scriptscriptstyle(1)}$ on the inverse intraspecies scattering length $1/a_{\scriptscriptstyle\textsl{HH}}$. All quantities are scaled by the intraspecies vdW length. }
   \label{HH}
  \end{figure}

For large $a_{\scriptscriptstyle\textsl{HH}}$, we find a general linear dependence of $a^{\scriptscriptstyle(1)}_{\scriptscriptstyle-}$ on the heavy-heavy interaction for all of the investigated systems as shown in Fig.\;\ref{HH}. Figure.\;\ref{HH}(b) shows the lines of $a^{\scriptscriptstyle(1)}_{\scriptscriptstyle-}$ for $^{85}$Rb\,-$^{85}$Rb\,-$^{3}$He$^*(2^3S)$ and $^{172}$Yb\,-$^{172}$Yb\,-$^{6}$Li systems as a function of $1/a_{\scriptscriptstyle\textsl{HH}}$. Notably, the two systems follow the same line. Investigating the mass and interaction parameters of the two systems reported in Table\;\ref{t2}, we find that they have very similar ratios of mass and the vdW coefficient. In our previous work, we also observed a similar phenomenon in the $^{87}$Rb\,-$^{87}$Rb\,-$^{3}$He$^*(2^3S)$ and $^{174}$Yb\,-$^{174}$Yb\,-$^{6}$Li systems\;\cite{Wumengshan2016}.

 Figure.\;\ref{HH} intuitively shows that the intraspecies scattering length has different effects on the value of $a_{\scriptscriptstyle-}^{\scriptscriptstyle(1)}$ according to the mass ratio of the system, and the slope of every line can be used to measure the degree of influence. Considering different HHL systems, it is observed that the slope is inversely proportional to the mass ratio $M/m$ and also determined by the ratio $(\large\frac{C_{6,\scriptscriptstyle\textsl{HL}}}{C_{6,\scriptscriptstyle \textsl{HH}}}\large)$ at the same time( see Fig.\;\ref{HH}(b)). These points have never been discussed before and are valuable implication to the universal behavior of 3BP in heteronuclear systems. Figure.\;\ref{HH} lays the basis for our understanding of the universal behavior of $a_{\scriptscriptstyle-}^{\scriptscriptstyle(1)}$ and allows us to extract its form from numerical results.
\begin{table}
\caption{\label{t2} The mass, two-body vdW dispersion coefficients, and slope of the ten HHL systems. The column `S' represents the slope, and `C' is the constant in our fits.}
\renewcommand\arraystretch{0.7}
\begin{tabular*}{16.7cm}{@{\extracolsep{\fill}}lccccccccccc}
   \hline
   \hline
$system( {\scriptstyle \textsl{H-H-L}})$&$M$&$m$ &$M/m$ &$C_{\scriptscriptstyle\textsl{6,HH}}$&$C_{\scriptscriptstyle\textsl{6,HL}}$&$\frac{C_{\scriptscriptstyle\textsl{6,HH}}}{C_{\scriptscriptstyle\textsl{6,HL}}}$ &$r_{\scriptscriptstyle\textsl{vdW,HH}}$ &$r_{\scriptscriptstyle\textsl{vdW,HL}}$&$C $ &$S$ \\
   \hline
$^{85}$Rb\,-$^{85}$Rb\,-$^{41}$K          &$85.4678$&$40.962$ &$2.087$ &$4698.0^a$ &$4106.5^b$ &$1.144$ &$82.24$ &$71.35$  &$8.757$ &$35.94$ \\
$^{85}$Rb\,-$^{85}$Rb\,-$^{40}$K          &$85.4678$&$39.964$ &$2.139$ &$4698.0^a$ &$4106.5^b$ &$1.144$ &$82.24$ &$71.05$  &$8.693$&$34.55$\\
$^{85}$Rb\;-$^{85}$Rb\,-$^{23}$Na         &$85.4678$&$22.99 $ &$3.718$ &$4698.0^a$ &$2581.5^b$ &$1.820$ &$82.24$ &$57.14$  &$8.537$&$17.74$ \\
$^{85}$Rb\,-$^{85}$Rb\,-$^{7}$Li          &$85.4678$&$7.016 $ &$12.182$&$4698.0^a$ &$2468.3^b$ &$1.903$ &$82.24$ &$43.70$  &$8.446$ &$5.26$ \\
$^{85}$Rb\,-$^{85}$Rb\,-$^{6}$Li          &$85.4678$&$6.015$  &$14.209$&$4698.0^a$ &$2468.3^b$ &$1.903$ &$82.24$ &$42.16$  &$8.567$&$4.64$ \\
$^{85}$Rb\,-$^{85}$Rb\,-$^{4}$He$^*(2^3S)$&$85.4678$&$4.003$  &$21.353$&$4698.0^a$ &$3832.0^c$ &$1.226$ &$82.24$ &$42.75$  &$8.525$&$3.29$ \\
$^{133}$Cs\,-$^{133}$Cs\,-$^{6}$Li        &$132.905$&$6.015$  &$22.095$&$6851.0^d$ &$3065.0^e$ &$2.235$ &$100.92$&$44.77$  &$8.735$&$3.02$ \\
$^{85}$Rb\,-$^{85}$Rb\,-$^{3}$He$^*(2^3S)$&$85.4678$&$3.016$  &$28.338$&$4698.0^a$ &$3832.0^c$ &$1.226$ &$82.24$ &$39.94$  &$8.629$&$2.54$ \\
$^{172}$Yb\,-$^{172}$Yb\,-$^{6}$Li        &$171.936$&$6.015$  &$28.584$&$1909.0^a$ &$1606.0^f$ &$1.189$ &$78.20$ &$38.19$  &$8.661$&$2.55$ \\
$^{176}$Yb\,-$^{176}$Yb\,-$^{6}$Li        &$175.943$&$6.015$  &$29.250$&$1909.0^a$ &$1606.0^f$ &$1.189$ &$78.65$ &$38.20$  &$8.710$&$2.52$ \\
\hline
\hline
\end{tabular*}
\begin{flushleft}
\vskip-1.5ex
$^{a}$From\cite{C6RbYbYbLi},
$^{b}$From\cite{C6RbKRbNaRbLi},
$^{c}$From\cite{C6RbHe},
$^{d}$From\cite{C6Cs},
$^{e}$From\cite{C6CsLi},
$^{f}$From\cite{C6RbYbYbLi}.
\end{flushleft}
\end{table}

 Table\;\ref{t2} summarizes the mass, two-body vdW dispersion coefficients, and ratios of $ M/m$ and $\large\frac{C_{6,\scriptscriptstyle\textsl{HL}}}{C_{6,\scriptscriptstyle \textsl{HH}}}\large$ for ten HHL systems. The last column is the slope labeled by `S' for every line in Figure.\;\ref{HH} fitting from our numerical results. It can be observed that the isotopic systems have different slopes though they possess the same $\large\frac{C_{6,\scriptscriptstyle\textsl{HL}}}{C_{6,\scriptscriptstyle \textsl{HH}}}\large$ ratio, which means that the mass ratio also affects the slope. $^{85}$Rb\,-$^{85}$Rb\,-$^{3}$He$^*(2^3S)$ and $^{172}$Yb\,-$^{172}$Yb\,-$^{6}$Li provide such an example, as they have similar ratios of $ M/m$ and $\large\frac{C_{6,\scriptscriptstyle\textsl{HL}}}{C_{6,\scriptscriptstyle \textsl{HH}}}\large$ and thus have the same slope. Fitting our numerical results, we find that the slope of every line can be written as  $ S \sim C^{2}(\frac{m}{M})(\frac{C_{\scriptscriptstyle\textsl{6,HL}}}{C_{\scriptscriptstyle
 \textsl{6,HH}}}\large)^{\frac{1}{6}} $, where C is a constant. Its values for every HHL system are listed in Table\;\ref{t2}. We find that they change a little for different systems and are about $8.7$ for all the investigated HHL system. Since the ratio $\large\frac{C_{6,\scriptscriptstyle\textsl{HL}}}{C_{6,\scriptscriptstyle \textsl{HH}}}\large$ is a similar value for different systems, as reported in Table\;\ref{t2}, the mass ratio $\frac{m}{M}$ is thus the main factor contributing to the slope term. This could explain why the 3BP of large mass ratio systems is insensitive to the variation in the intraspecies scattering length.

\begin{figure}
\centering
\includegraphics[width=0.69\linewidth]{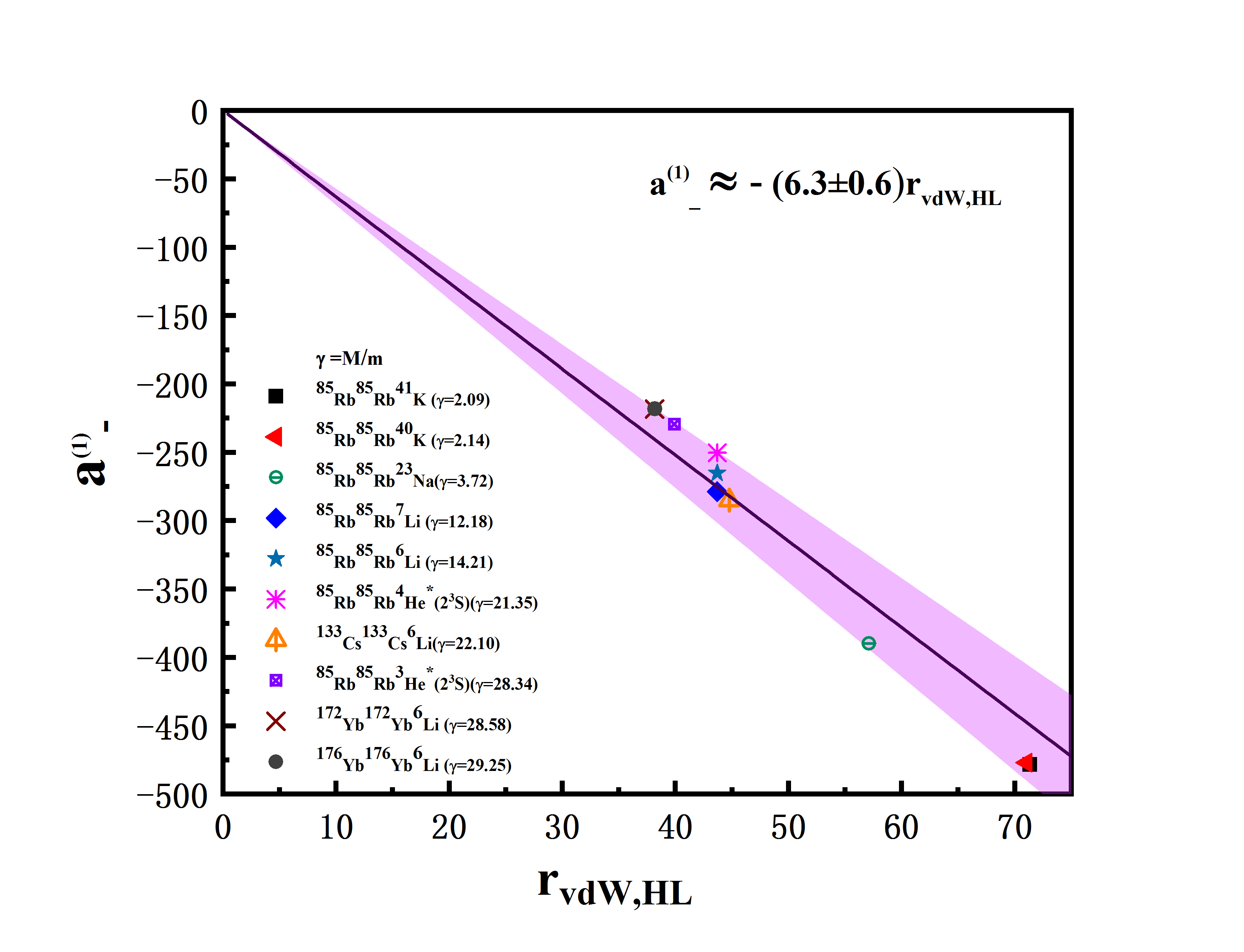}
\caption{The vdW universality of the 3BP in ten HHL systems with the resonant heavy-heavy interaction. The solid line along with the shaded area is plotted using the formula $a^{\scriptscriptstyle(1)}_{\scriptscriptstyle-} \approx -(6.3 \pm 0.6) r_{\scriptscriptstyle\textsl{vdW,HL}}$. The points are our numerical values.}
\label{HL}
\end{figure}

 When the intraspecies scattering length approaches infinity, we find the $a^{\scriptscriptstyle(1)}_{\scriptscriptstyle-}$ of these systems all lie in a universal range. Through fitting our numerical results, a universal expression is given: $a^{\scriptscriptstyle(1)}_{\scriptscriptstyle-} \approx -(6.3 \pm 0.6) r_{\scriptscriptstyle \textsl{vdW,HL}}$. The average value $-6.3\,r_{\scriptscriptstyle\textsl{vdW,HL}}$ is obtained by fitting our numerical values, and the uncertainty indicates the largest deviations from the average. Figure.\;\ref{HL} shows a comparison between our calculated $a^{\scriptscriptstyle(1)}_{\scriptscriptstyle-}$ values for the resonant heavy-heavy interaction and the universal constant, where all data for these ten systems fall into the universal range.

\begin{figure}
\centering
\includegraphics[width=0.9\linewidth]{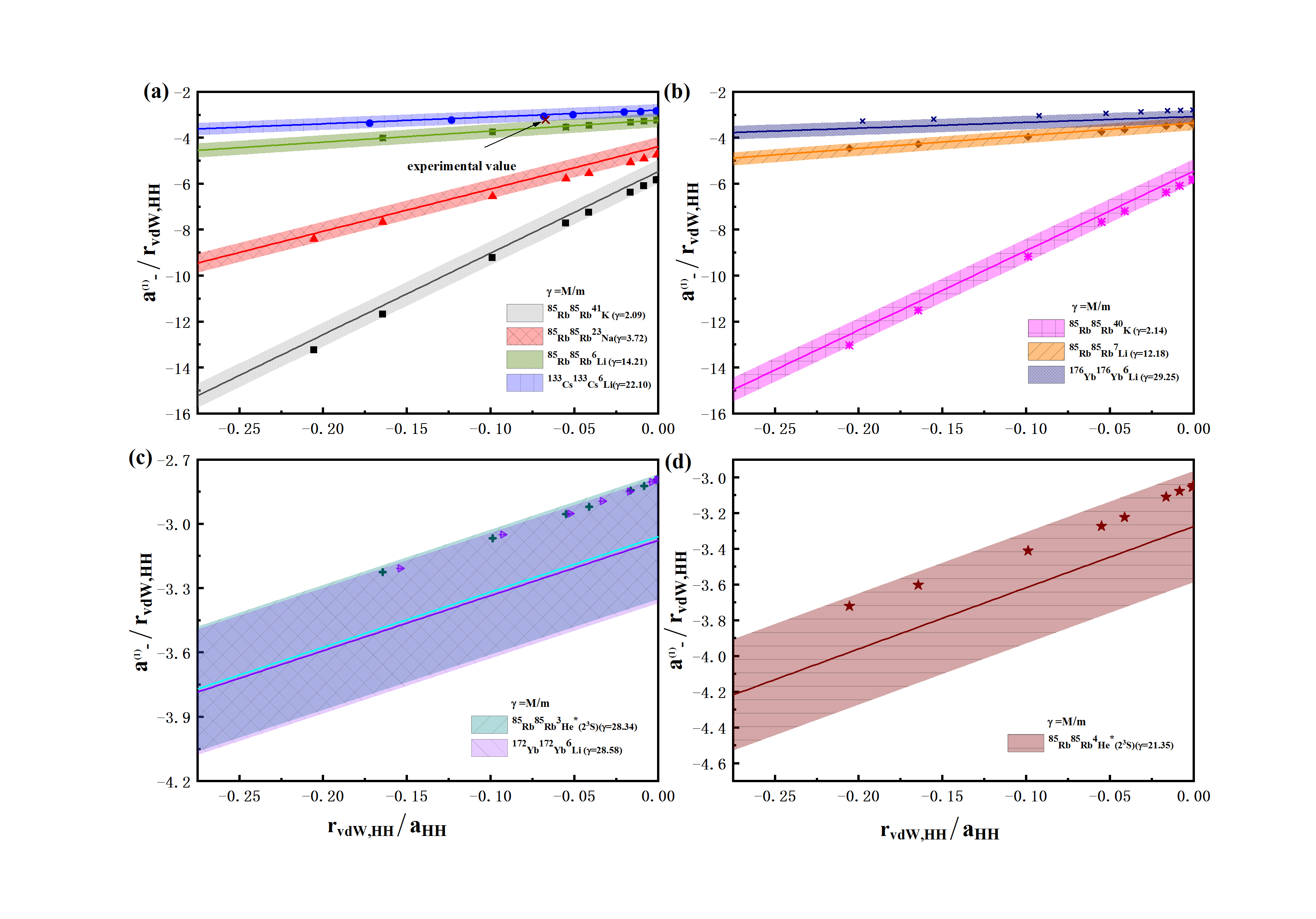}
\caption{Comparison of the numerical values (points) and the analytic form (\ref{eq:3BP}) (line) for different HHL systems are shown in (a) (b) (c) (d) respectively. The shaded region specifies a band in which there is a $10\%$ deviation from the exact value of $-6.3\,r_{\scriptscriptstyle\textsl{vdW}}$. The symbol $\times$ in (a) indicates the experimental value\;\cite{Tung2014cslia-}.}
\label{Slope}
\end{figure}

After we obtain the universal expression of slope and the $a_{\scriptscriptstyle-}^{\scriptscriptstyle(1)}$ at the intraspecies resonant interaction, we can write the $a_{\scriptscriptstyle-}^{\scriptscriptstyle(1)}$ as a function of $a_{\scriptscriptstyle \textsl{HH}}$  in the following form:
\begin{align}
\label{eq:3BP}
 a^{\scriptscriptstyle(1)}_{\scriptscriptstyle-}/r_{\scriptscriptstyle\textsl{vdW,HH}} \approx 8.7^{2}(\frac{m}{M})(\frac{C_{\scriptscriptstyle\textsl{6,HL}}}{C_{\scriptscriptstyle
 \textsl{6,HH}}}\large)^{\frac{1}{6}}\frac{r_{\scriptscriptstyle\textsl{vdW,HH}}}
 {a_{\scriptscriptstyle\textsl{HH}}} - (6.3 \pm 0.6)  \frac{r_{\scriptscriptstyle\textsl{vdW,HL}}}{r_{\scriptscriptstyle\textsl{vdW,HH}}}\,.
\end{align}
Equation. (\ref{eq:3BP}) can be used to estimate the $a^{\scriptscriptstyle (1)}_{-}$ value given the value of the background scattering length $a_{\scriptscriptstyle \textsl{HH}}$ for HHL systems whose mass ratio ranges from $2$ to $30$. The value of the background scattering length $| a_{\scriptscriptstyle \textsl{HH}} |$ should be greater than $4\, r_{\scriptscriptstyle\textsl{vdW,HH}}$. As shown in Fig.\;\ref{Slope}, Eq. (\ref{eq:3BP}) gives a reasonable estimate of $a^{\scriptscriptstyle (1)}_{-}$ compared to the exact numerical results. The solid line with the shaded region is given by Eq. (\ref{eq:3BP}), and the points are our numerical values. The shaded region comes from the uncertainty of $a^{\scriptscriptstyle (1)}_{-}$ when two heavy atoms are in resonant interaction. It is worth mentioning that the first Efimov resonance in the $^{133}$Cs\,-$^{133}$Cs\,-$^{6}$Li system with a negative background scattering length has been observed; the experimental value is $a^{\scriptscriptstyle(1)}_{\scriptscriptstyle-} = -323(8)\, a_0 = -7.2\, r_{\scriptscriptstyle\textsl{vdW,CsLi}} $\;\cite{Tung2014cslia-}. We show the position in Fig.\;\ref{Slope}(a) with the symbol $\times$, which is in the range predicted by the analytic formula (\ref{eq:3BP}).

\section{Conclusions}
In summary, we have investigated $a^{\scriptscriptstyle(1)}_{\scriptscriptstyle-}$ for ten mass-imbalanced systems. Our calculations provide numerical evidence that systems with negative intraspecies scattering lengths are more suitable for observing Efimov ground states than those with positive ones. Our results show that an experiment may observe Efimov effects in $^{85}$Rb\,-$^{85}$Rb\,-$^{41,40,39}$K systems under current experimental conditions. Investigating the dependence of the $a^{\scriptscriptstyle(1)}_{\scriptscriptstyle-}$ on the intraspecies interaction in ten mass-imbalanced systems, we find that $a^{\scriptscriptstyle(1)}_{\scriptscriptstyle-}$ has a dramatically strong dependence on the background scattering length $a_{\scriptscriptstyle\textsl{HH}}$ for the less mass-imbalanced systems, while for strongly mass-imbalanced systems, the intraspecies interaction plays less of an important role in determining the value of $a^{\scriptscriptstyle(1)}_{\scriptscriptstyle-}$. The point that the intraspecies scattering length has different effects in heteronuclear systems according to their mass ratio is discussed. For large  $a_{\scriptscriptstyle\textsl{HH}}$, a linear fit of $a^{\scriptscriptstyle (1)}_{-}$ as a function of the background scattering length is obtained for different mass ratios, and generally shows how the mass ratio $M/m$ and the vdW interaction affect the universal behavior of the 3BP. For an HHL system whose mass ratio ranges from $2$ to $30$, the formula can estimate the $a^{\scriptscriptstyle(1)}_{\scriptscriptstyle-}$ given the value of the background scattering length with an approximately $10\%$ deviation. Note that there are fewer experimental values of the 3BP for the negative intraspecies scattering length case, and the unique experimental result of $a^{\scriptscriptstyle(1)}_{\scriptscriptstyle-}$ with a negative scattering length in $^{133}$Cs\,-$^{133}$Cs\,-$^{6}$Li is consistent with the universal position predicted by the analytical formula. Our results also show that $a^{\scriptscriptstyle (1)}_{-}$ is nearly a constant expressed in terms of the vdW length $r_{\scriptscriptstyle \textsl{vdW,HL}}$, $a^{\scriptscriptstyle(1)}_{\scriptscriptstyle-} = -(6.3 \pm 0.6) r_{\scriptscriptstyle\textsl{vdW,HL}}$, with the resonant heavy-heavy interaction. Finally, we should note that the present results are valid for broad Feshbach resonances but less so for narrow ones due to the single-channel treatment of the atomic interaction.

We thank C.H. Greene, Peng Zhang, and Chao Gao for helpful discussions. H.-L. Han was
supported by the National Natural Science Foundation of China under Grants
No. 11874391 and No. 11634013 and the National Key Research and Development
Program of China under Grant No. 2016YFA0301503. M.-S. Wu was supported by the National Natural Science Foundation of China under Grant
No. 11704399. T.-Y. Shi was supported by the Strategic Priority Research Program of the Chinese Academy of Sciences under Grant No. XDB21030300.


%

\end{document}